\documentclass[a4paper,fleqn]{cas-dc}

\usepackage[square, numbers]{natbib}
\usepackage{hyperref}
\hypersetup{colorlinks=true, linkcolor=blue, citecolor=blue}
\usepackage{booktabs}
\usepackage[]{mdframed}
\usepackage{amssymb}
\usepackage{setspace}
\usepackage[utf8]{inputenc}
\usepackage{moresize}
\usepackage{tikz}

\usepackage{nomencl}

\usepackage{etoolbox}
\renewcommand\nomgroup[1]{%
  \item[\bfseries
  \ifstrequal{#1}{A}{}{%
  }%
]}

\makenomenclature

\def\tsc#1{\csdef{#1}{\textsc{\lowercase{#1}}\xspace}}
\tsc{WGM}
\tsc{QE}

\begin{document}
\let\WriteBookmarks\relax
\def\floatpagepagefraction{1}
\def\textpagefraction{.001}

\shorttitle{Ultrafast transfer of low-mass payloads to Mars and beyond using aerographite solar sails}    

\shortauthors{Karlapp~et~al.}  

\title [mode = title]{Ultrafast transfer of low-mass payloads to Mars and beyond using aerographite solar sails}

\author[1]{Julius Karlapp~}

\cormark[1]

\affiliation[1]{organization={Institute of Aerospace Engineering, Technische Universit\"at Dresden},
    addressline={Marschnerstr. 32}, 
    city={01307 Dresden},
    country={Germany}}

\author[2]{René Heller~}

\fnmark[1]

\affiliation[2]{organization={Max Planck Institute for Solar System Research},
    addressline={Justus-von-Liebig-Weg 3}, 
    city={37077 Göttingen},
    country={Germany}}

\author[1]{Martin Tajmar~}

\cortext[1]{
                ~Corresponding author ~(E-Mail: \textcolor{blue}{~julius.karlapp@gmail.com}).
            }

\fntext[1]{
                ~Presenting author, 8th Interstellar Symposium 2023.
}

\begin{abstract}
With interstellar mission concepts now being under study by various space agencies and institutions, a feasible and worthy interstellar precursor mission concept will be key to the success of the long shot. Here we investigate interstellar-bound trajectories of solar sails made of the ultra lightweight material aerographite. Due to its extremely low density ($0.18\,{\rm kg\,m}^{-3}$) and high absorptivity ($\mathcal{A}{\sim}1$), a thin shell can pick up an enormous acceleration from the solar irradiation. Payloads of up to 1\,kg can be transported rapidly throughout the solar system, e.g. to Mars and beyond. Our simulations consider various launch scenarios from a polar orbit around Earth including directly outbound launches as well as Sun diver launches towards the Sun with subsequent outward acceleration. We use the {\tt poliastro} Python library for astrodynamic calculations. For a spacecraft with a total mass of 1\,kg (including 720\,g aerographite) and a cross-sectional area of $10^4\,{\rm m}^2$, corresponding to a shell with a radius of 56\,m, we calculate the positions, velocities, and accelerations based on the combination of gravitational and radiation forces on the sail. We find that the direct outward transfer to Mars near opposition to Earth results in a relative velocity of $65\,{\rm km\,s}^{-1}$ with a minimum required transfer time of 26\,d. Using an inward transfer with solar sail deployment at 0.6\,AU from the Sun, the sail's velocity relative to Mars is $118\,{\rm km\,s}^{-1}$ with a transfer time of 126\,d, where Mars is required to be in one of the nodes of the two orbital planes upon sail arrival. Transfer times and relative velocities can vary substantially depending on the constellation between Earth and Mars and the requirements on the injection trajectory to the Sun diving orbit. The direct interstellar trajectory has a final velocity of $109\,{\rm km\,s}^{-1}$. Assuming a distance to the heliopause of 120\,AU, the spacecraft reaches interstellar space after 5.3\,yr. When using an initial Sun dive to 0.6\,AU instead, the solar sail obtains an escape velocity of $148\,{\rm km\,s}^{-1}$ from the solar system with a transfer time of 4.2\,yr to the heliopause. Values may differ depending on the rapidity of the Sun dive and the minimum distance to the Sun. The mission concepts presented in this paper are extensions of the 0.5\,kg tip mass and $196\,{\rm m}^2$ design of the successful IKAROS mission to Venus towards an interstellar solar sail mission. They allow fast flybys at Mars and into the deep solar system. For delivery (rather than fly-by) missions of a sub-kg payload the biggest obstacle remains in the deceleration upon arrival.
\end{abstract}


\maketitle


\section{Introduction}\label{Introduction}
\noindent

From the first medium distance ride of a steam locomotive in 1804, the `Penydarren' of British mining engineer Richard Trevithick, at about $4\,{\rm km\,hr}^{-1}$ to the current heliocentric escape velocity of the Voyager 1 probe at $17\,{\rm km\,s}^{-1}$, humanity has managed a velocity gain of their mechanical and automated vehicles of over four orders of magnitude within about 200\,yr \citep{2017MNRAS.470.3664H}. If technological progress continues to increase humanity's top velocitys at this rate, that is, a doubling every 15\,yr for about another seven doublings, we would reach 1\,\% of the velocity of light ($c$) within little more than 100\,yr from now. At this point, our travel to the stars might become reality. With our closest stellar neighbor Proxima Centauri located at a distance of about 1.3\,pc (4.2\,ly) \cite{1999AJ....118.1086B} a velocity of $0.1\,c$ implies a travel time of about 42\,yr, well within a human lifetime and within the 46\,yr of operation time of the Voyager\,1 mission.

Recent studies suggest that the technology and knowledge components for interstellar missions are in reach, such as miniaturized electronic devices \citep{Duduta2018}, advanced information processing \cite{2016JBIS...69..278M,2019IJAsB..18..267H}, ultra-light materials such as graphene \cite{2004Sci...306..666N} and aerographite \cite{Mecklenburg2012} that could be used for light sails, and predictions of the light sail dynamics \cite{Cassenti1997,Dachwald2005,2020A&A...641A..45H}. Solar sail technology has been demonstrated to work during various missions in the solar system, such as IKAROS \cite{2013AcAau..82..183T}, NanoSail-D\footnote{\href{https://ntrs.nasa.gov/citations/20110015650}{~https://ntrs.nasa.gov/citations/20110015650}}, and LightSail-2 \cite{Mansell2020} (see \cite{2020A&A...641A..45H,macdonald2014advances} for further references).

An intermediate step toward interstellar travel are interstellar precursor missions \cite{2013JBIS...66..252F}, some of which have been proposed for a comet rendevous \cite{2004JSpRo..41..140H} and deep solar system exploration, e.g. the search for the suspected Planet Nine \cite{2020A&A...641A..45H,2020ApJ...895L..35H}. 

In this paper, we explore an interstellar precursor scenario that we expect to be more relevant in the near future. Driven by NASA's goal to put humans on Mars by the end of the 2030s \cite{Linck2019} and related plans by the SpaceX company, we explore the potential of solar lightsails to carry small payloads to Mars in short times, that is, weeks to months. The energy and thus monetary demands of conventional chemical rocketry depend strongly on the Earth-Mars orbital configuration. As a result, launches are usually chosen to follow a low-energy Hohmann transfer orbit, the launch window of which opens up every 780\,d or 2.1\,yr.  


\begin{mdframed}[leftmargin=0.02cm, innertopmargin=0.1cm, innerbottommargin=1.05cm]

        \nomenclature[A]{\(\mathcal{A}\)}{Absorptivity}
        \nomenclature[A]{\(c\)}{Speed of light}
        \nomenclature[A]{\(\kappa_{\rm rad}\)}{Radiation coupling constant}
        \nomenclature[A]{\(F_{\mathrm{tot}}\)}{Force applied to the sail}
        \nomenclature[A]{\(F_{\mathrm{rad}}\)}{Force of solar radiation}
        \nomenclature[A]{\(F_{\mathrm{grav}}\)}{Gravitational force}
        \nomenclature[A]{\(r\)}{Solar distance}
        \nomenclature[A]{\(L_\odot\)}{Solar luminosity}
        \nomenclature[A]{\(S\)}{Sail cross sectional area}
        \nomenclature[A]{\(M_\odot\)}{Solar mass}
        \nomenclature[A]{\(m\)}{Spacecraft mass}
        \nomenclature[A]{\(l\)}{Sail radius}
        \nomenclature[A]{\(m_{\mathrm{aero}}\)}{Aerographite mass}
        \nomenclature[A]{\(\rho\)}{Aerographite density}
        \nomenclature[A]{\(\varepsilon\)}{Shell thickness}
        \nomenclature[A]{\(r_{\mathrm{Hill}}\)}{Hill radius}
        \nomenclature[A]{\(a\)}{Semi-major axis}
        \nomenclature[A]{\(M_{\rm p}\)}{Planets mass}
        \nomenclature[A]{\(r_{\mathrm{Hill,Earth}}\)}{Hill radius of Earth}
        \nomenclature[A]{\(r_{\mathrm{Hill,Mars}}\)}{Hill Radius of Mars}
        \nomenclature[A]{\(i\)}{Inclination}
        \nomenclature[A]{\(h_{\rm minimum}\)}{Minimum orbit height}
        \nomenclature[A]{\(v_{\rm exit,Earth}\)}{Exit velocity from Earth's sphere of influence}
        \nomenclature[A]{\(t_{\rm exit,Earth}\)}{Exit time from Earth's sphere of influence}
        \nomenclature[A]{\(\Delta v\)}{Change in velocity}
        \nomenclature[A]{\(t\)}{time}
        \nomenclature[A]{\(v\)}{Velocity}
        \nomenclature[A]{\(a\)}{Acceleration}
        \nomenclature[A]{\(t_{\rm entry}\)}{Entry time}
        \nomenclature[A]{\(t_{\rm Mars}\)}{Duration inside martian sphere of influence}
        \nomenclature[A]{\(t_{\rm end}\)}{Final transfer time}
        \nomenclature[A]{\(v_{\rm end}\)}{Final transfer velocity}
        \nomenclature[A]{\(a_{\rm max}\)}{Maximum acceleration}
        \nomenclature[A]{\(v_{\rm entry}\)}{Entry velocity}
        \nomenclature[A]{\(t_{\rm deploy}\)}{Time sail deployment}
        \nomenclature[A]{\(\Delta t\)}{Duration}
        \nomenclature[A]{\(v_{\rm Earth}\)}{Orbital velocity of Earth around the Sun}
        \nomenclature[A]{\(R_{\oplus}\)}{Radius reference body}
        \nomenclature[A]{\(g\)}{Earth surface gravity}

        \onehalfspacing
        \printnomenclature[1.35cm]
\end{mdframed}  

\noindent
These time scales have been proven manageable for fully robotic missions and even for more longterm missions to Mars like the Mars sample return, which are effectively a sequence of Mars missions that build on each other. Once the human factor comes into play, however, risk and emergency assessments change substantially and a wait time of months to years, e.g. for medical supply or replenishment of essential materials or devices may be too long.

In this study we feed, for the first time, results from laboratory studies on the material properties of aerographite \cite{Mecklenburg2012} into solar sail trajectory simulations that aim at Mars and beyond. Aerographite, with its extremely low density of $0.18\,{\rm kg\,m}^{-3}$ (compared to $2700\,{\rm kg\,m}^3$ for aluminum foil, for example) and a radiation coupling constant $\kappa_{\rm rad}~\sim~1$ that permits very efficient transformation of solar irradiation to acceleration, stands out as a material for solar sail applications \cite{2020A&A...641A..45H}. Being a carbon-based aerogel, aerographite can be described as a foam-like material with a microscopic structure composed of $\mu$m-sized carbon microtubes \cite{schuchardt2014mechanical, chandrasekaran2017carbon, hirahara2017ultra}. Important for the structural integrity of any macroscopic application, aerographite is characterized by a high elasticity. Elastic deformation occurs up to a compression rate of $95\%$. The combination of low density, efficient radiative coupling, and high elasticity makes aerographite an excellent candidate material for solar sails.

Beyond solar sail concepts to Mars, we study the dynamical properties of aerographite sails that reach escape velocities from the solar system. Hence, solar sail missions to Mars can serve as natural precursors for interstellar missions to establish the technologies for long-distance communication, deep space navigation etc.

\section{Methods - Simulation Script}\label{Methods}

\noindent
To investigate interplanetary and interstellar transfers and to simulate possible trajectory progressions we use the Python programming language in combination with the {\tt poliastro} library for astrodynamic calculations and visualization \footnote{\href{https://docs.poliastro.space/en/stable/}{~https://docs.poliastro.space/en/stable/}}. Our objective is to use numerical calculations of the positions, velocities, and accelerations for various sail trajectories using adaptive time steps.

The force on our hypothetical sail is parameterized as

\begin{equation} \label{thrustfunction}
    F_{\mathrm{tot}}=F_{\mathrm{rad}}+F_{\mathrm{grav}}=\frac{1}{r^2}\Bigl(\frac{L_\odot}{4 \pi c} S \kappa_{\mathrm{rad}} \ - \ G M m \Bigr) \ ,
\end{equation}

\noindent
where $r$ is the solar distance, $L_\odot$ the solar luminosity, $S$ the sail cross sectional area, $M$ is the mass of the primary source of gravity for the sail that defines the current sphere of influence, and $m$ the spacecraft mass \cite{2020A&A...641A..45H}.

We consider an aerographite sail with $\kappa_{\mathrm{rad}}=1$ that is shaped as a hollow sphere with a cross-sectional area of $S=(100\,{\rm m})^2=10^4\,{\rm m}^2$. This corresponds to a sail radius of $l=\sqrt{S/\pi}=56.4$\,m. We assume spherically symmetric distribution of the sail mass with a total mass of $m=1$\,kg. Although our equations are ignorant as to the contribution of the aerographite and the payload to the total mass of the spacecraft, we can use Eq. (11) from Heller et al. \cite{2020A&A...641A..45H} to estimate the aerographite mass $m_{\rm aero}=\pi l^2 4\rho\varepsilon$, where $\varepsilon$ is the shell thickness. For a nominal 0.1\,mm thin hollow sphere with a radius of 56.4\,m this gives $m_{\rm aero}=720$\,g, leaving 280\,g for payload. Using a hollow aerographite hemisphere would halve the aerographite mass and permit 640\,g of payload.

During interplanetary transfer we assume a two-body problem of the sail and its dominant gravitational counterpart, the latter of which can be the Sun, the Earth or Mars in our case. As long as the sail is beyond the sphere of influence (SOI) of a planet, we apply Eq.~\eqref{thrustfunction} with $M = M_\odot$, or else we substitute the solar mass with the mass of the respective planet. We estimate the planetary SOI via the Hill radius

\begin{equation} \label{hillradius}
    r_{\mathrm{Hill}}=a\cdot\sqrt[3]{\frac{M_{\rm p}}{3M_\odot}} ~~,
\end{equation}

\noindent
inside of which the gravitational attraction of the planet on a point mass (i.e. the sail) dominates over the gravitational force from the Sun. The Hill radius is defined by the solar mass, the mass of the planet $M_{\rm p}$, and the semi-major axis ($a$) of the Sun-planet orbit, which is approximated to be circular. The Earth's Hill radius is $r_{\mathrm{Hill, Earth}}~\approx~1.5\,{\times}\,10^6$\,km, while the martian Hill radius is $r_{\mathrm{Hill, Mars}}~\approx~1.1\,{\times}\,10^6$\,km \cite{gurfil2016celestial, deIacoVeris2018practical}.

\section{Results}\label{Results}

\noindent
In \hyperref[scenarios]{Table~\ref*{scenarios}} we summarize our setup of the possible launch scenarios. We consider both direct outward transfers and inward transfers. The outward injection calculates for a launch from a parking orbit around Earth radially away from the Sun (initially defined in the gravitational sphere of influence of Earth), while the inward injection calculates for an initial conventional launch towards the Sun with a delayed sail deployment in a lower solar orbit (Sun diver approach, initially defined in the solar gravitational sphere of influence). 

\subsection{Outward Trajectories}

\noindent
We now investigate the ability of our hypothetical spacecraft to reach Mars or interstellar space by calculating the spacecraft position, velocity, and acceleration as a function of time.

\begin{table}[h]
    \rmfamily
    \caption{\rmfamily General orbit configurations for heliocentric injections of a solar sail spacecraft ($m=1\,{\rm kg}$, $A=10^4\,{\rm m}^2$). All outward injections are referring to the geocentric reference system (GCRS) because the parking orbit is defined around Earth, whilst the inward injections are defined in the heliocentric reference system (HCRS) because the spacecraft is simulated directly in the solar gravitational sphere.}\label{scenarios}
    \resizebox{\columnwidth}{!}{%
    \begin{tabular}{lllll}                                          
        \toprule
        Heliocentric Injection &\multicolumn{2}{l}{\textbf{Outward (GCRS)}}   &\multicolumn{2}{l}{\textbf{Inward (HCRS)}}\\
        \midrule
        Destination             & Mars     & Interstellar       & Mars              & Interstellar \\
        \midrule
        Orbit                   & Polar    & Polar              & Solar             & Solar \\
        semi-major axis $[km]$  & $42000$  & $42000$            & $149.6\cdot10^6$  & $149.6\cdot10^6$ \\
        Inclination $[^\circ]$  & $113.4$  & $113.4$            & $0$               & $0$ \\
        Eccentricity $[-]$      & $0.24$   & $0.24$             & $0.0167$          & $0.0167$ \\
    \bottomrule
    \end{tabular}%
    }
\end{table}

\subsubsection{Outward Mars Transfer} \label{OutwardMarsTransfer}

\noindent
For a direct outwards transfer from Earth to Mars, Mars needs to be near opposition since the vector of the solar radiation pressure will push the spacecraft radially away from the Sun. Therefore two possible starting orbits (parking orbits) around Earth are conceivable, a High Elliptical Orbit (HEO) or a Polar Orbit (PO) \cite{karlapp_julius_2023_8082084}. Previous investigations showed that other standard orbits tend to produce collisions with Earth or erratic behavior of the spacecraft \citep[][Sect.~3 therein]{2020A&A...641A..45H}.

\begin{figure}[t]
	\centering
		\input{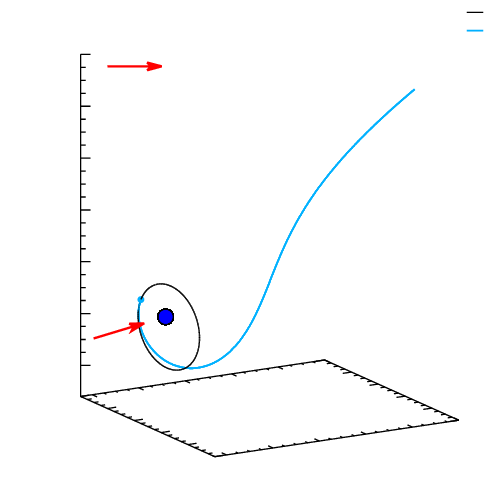}
	  \caption{The polar orbit (PO, black) with the escaping trajectory (light blue) for a direct outward ejection. The PO is in the $y$-$z$ plane. The solar illumination (red arrow) is along the $x$ axis.}
        \label{shotOutMarsSOIEarth}
\end{figure}

For an interplanetary transfer mission the use of a polar orbit is preferable (\hyperref[scenarios]{Table~\ref*{scenarios}}). Due to the orientation of the polar orbit the orbital plane of the spacecraft is oriented perpendicular to the solar radiation vector as shown in \hyperref[shotOutMarsSOIEarth]{Fig.~\ref*{shotOutMarsSOIEarth}}. Therefore every point along the orbit experiences a continuous irradiation from the sun (the Earth's shadow does not matter). A time independent deployment of the sail is possible. Referring to the geocentric reference system the inclination angle of the orbit plane is $113.4^\circ$ due to the tilt of the Earth's spin axis, whereas in the heliocentric reference system the inclination is $90^\circ$. The semi-major axis of the PO is chosen to be $42000\,{\rm km}$. For smaller values the produced trajectories wont overcome the gravitational pull of the Earth. For larger values the gain in net acceleration is marginal. Since the altitude of the PO around the Earth is negligible compared to the scale of an astronomical unit (PO altitude is $2.4\,{\times}\,10^{-4}$\,AU) the magnitude of the solar radiation pressure can be assumed to be constant.

\hyperref[shotOutMarsSOIEarth]{Figure~\ref*{shotOutMarsSOIEarth}} shows the PO (black) around Earth (blue), chosen to calculate the outward injections following \hyperref[scenarios]{Table~\ref*{scenarios}}. The escaping  trajectory (light blue) can be seen to align with the vector of the solar radiation pressure (red arrow) over long distances.

Upon deployment of the sail (e.g. unfolding in space) in the PO, the trajectories initially spiral away from Earth due to the two acting forces of the solar radiation pressure and gravitational force of Earth. As the sail picks up speed, its trajectory resembles more and more a straight line. Any live correction of the trajectory would need additional attitude control mechanism, which we neglect in our simulations. Instead, we consider the sail a projectile under the combined photogravitational effects.

\begin{figure}[t]
	\centering
		\input{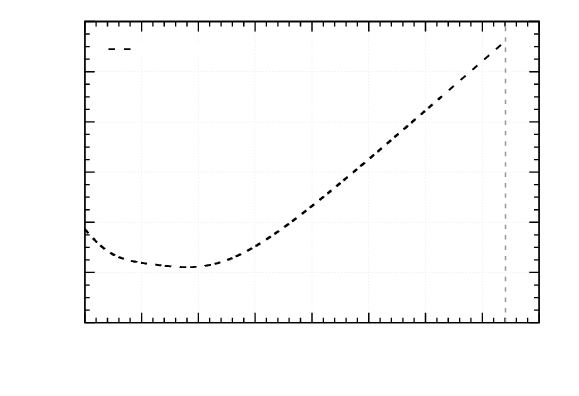}
	  \caption{Diagram of the spacecraft velocity $v$ as a function of time $t$ inside Earth's SOI for an outward transfer.}
        \label{vtDiagramOutMarsSOIEarth}
\end{figure}

The sail initially follows a curly trajectory as shown in \hyperref[shotOutMarsSOIEarth]{Fig.~\ref*{shotOutMarsSOIEarth}}. The Earth's gravitational pull reduces the departing velocity by about $1.5\,{\rm km\,s}^{-1}$ over the course of about 1\,d (\hyperref[vtDiagramOutMarsSOIEarth]{Fig.~\ref*{vtDiagramOutMarsSOIEarth}}). Thereafter, the force from the solar irradiation takes over and induces a net acceleration of the sail. At the time the sail enters interplanetary space after 3.54\,d the spacecraft has an exit velocity from the Earth's SOI of 
$11.2\,{\rm km\,s}^{-1}$. \footnote{The identity with the escape velocity from Earth is pure coincidence.}

\begin{figure}[t]
	\centering
		\input{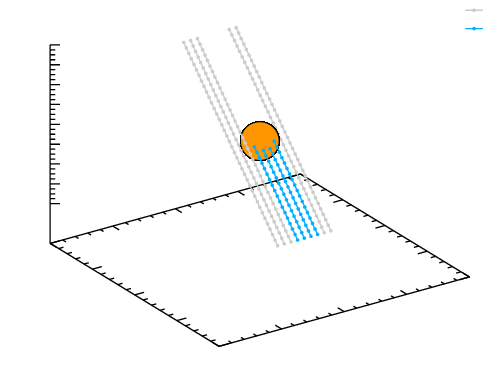}
	  \caption{Three-dimensional view of several trajectories inside the martian SOI. Trajectories in light blue color collide with Mars, trajectories in grey color denote fly-bys.}
        \label{shotOutMarsSOIMars}
\end{figure}

After the transition from the SOI of Earth to interplanetary space, that is, into the Sun's SOI the spacecraft enters the martian SOI. Inside the martian SOI several trajectories are presented (\hyperref[shotOutMarsSOIMars]{Fig.~\ref*{shotOutMarsSOIMars}}). Those trajectories only differ slightly in terms of their starting point from the initial parking orbit around Earth. For the purpose of a better visual inspection these trajectories are shown in the vicinity of Mars. Any trajectory that reaches an altitude of 200\,km or less above the martian surface is considered a collision (light blue) while the other trajectories (grey) escape the martian SOI.

\begin{figure}[t]
	\centering
		\input{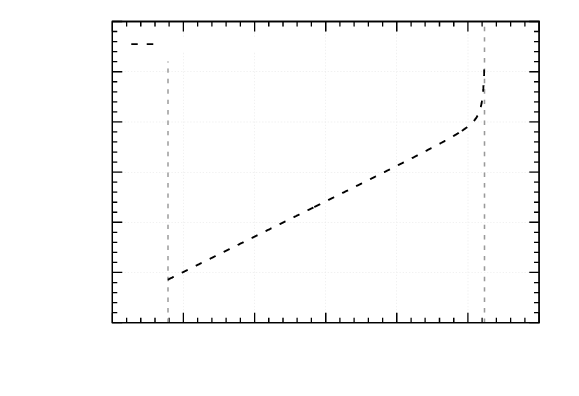}
	  \caption{Diagram of the spacecraft velocity $v$ as a function of time $t$ inside the martian SOI for an outward transfer.}
        \label{vt3DiagramOutMarsSOIMars}
\end{figure}

The spacecraft enters the martian SOI 
613.8\,hr or 
25.6\,{\rm d} after launch from the Earth's PO (\hyperref[vt3DiagramOutMarsSOIMars]{Fig.~\ref*{vt3DiagramOutMarsSOIMars}}). After 
just another 4.5\,hr inside the martian SOI, the spacecraft reaches Mars. 
During the flight inside the martian SOI the spacecraft is subject to an acceleration by the martian gravity and therefore gains 
$0.51\,{\rm km\,s}^{-1}$ which gives it a final velocity of 
$65.9\,{\rm km\,s}^{-1}$.

\subsubsection{Outward Interstellar Transfer} \label{OutwardInterstellarTransfer}

\noindent
Regarding the outward interstellar transfer we used the same polar parking orbit around Earth to deploy the spacecraft (\hyperref[scenarios]{Table~\ref*{scenarios}}). Instead of targeting Mars, however, all trajectories have been extended until they reached the boundary of the solar system, which we define at the distance of the heliopause at $r=120\,{\rm AU}$.

\begin{figure}[t]
	\centering
		\input{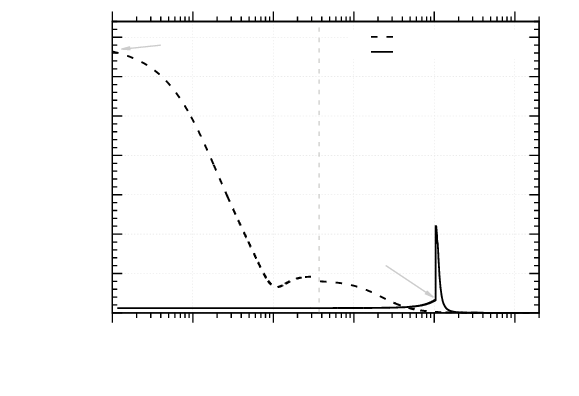}
	  \caption{Diagram of the spacecraft acceleration $a$ as a function of time $t$ for both outward and inward transfers.}
        \label{atDiagramOutInterstellar}
\end{figure}

The initial acceleration in the Earth's orbit corresponds to the maximum acceleration of 
$0.33\,{\rm m\,s}^{-2}$ (dashed line in \hyperref[atDiagramOutInterstellar]{Fig.~\ref*{atDiagramOutInterstellar}}), corresponding to about $3\,\%g$. The acceleration then decreases following the inverse square law as a function of distance from the Sun. The factor applies also for gravitational forces and was added to the general thrust equation (\hyperref[thrustfunction]{Eq.~\ref*{thrustfunction}}).

\begin{figure}[t]
	\centering
		\input{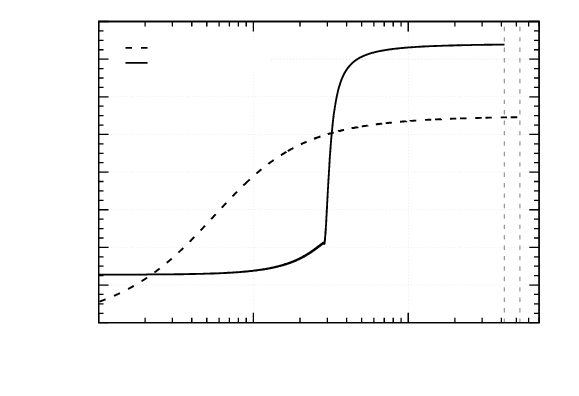}
	  \caption{Diagram of the spacecraft's velocity $v$ as a function of time $t$ in interplanetary space.}
        \label{vt2DiagramOutInterstellar}
\end{figure}

As a result, the velocity of the spacecraft increases continuously, but ever less with increasing distance to the Sun. The spacecraft ultimately reaches a terminal velocity of $v_{\mathrm{end}}=109\,{\rm km\,s}^{-1}$ after a duration of about $2\,{\rm yr}$ (\hyperref[vt2DiagramOutInterstellar]{Fig.~\ref*{vt2DiagramOutInterstellar}}). At this maximum speed, the spacecraft reaches the boundary of the solar system after $t_{\mathrm{end}}=5.3\,{\rm yr}$. 

We executed a limited series of trajectory simulations with variations of the launch configurations from the initial parking orbit around the Earth and found none that yielded any significant increase of the terminal velocity. The terminal velocity of the spacecraft can only be increased by reducing its overall mass or by increasing its cross sectional surface.

\subsection{Inward Trajectories}

\noindent
An alternative approach to increase the terminal velocity of the spacecraft is to reduce the distance to the Sun for the initial sail deployment. The increased photon pressure in the vicinity to the Sun results in a higher radiation force component and therefore a shorter transfer time than the direct outward approach from Earth.

\begin{figure}[t]
	\centering
		\input{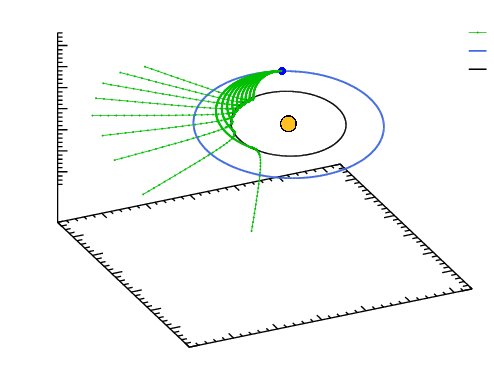}
	  \caption{The Sun diver approach allows the spacecraft to achieve several radial trajectory directions inside Earth's orbital plane around the Sun. The required breaking maneuver in the parking orbit around Earth is labeled $\Delta v_{\rm breaking}$ for each direction. Steeper maneuvers require a higher amount of $\Delta v$.}
        \label{sun-diver2d}
\end{figure}

To prove this thesis our script was modified to simulate a two phase flight. For the first phase a conventional impulse maneuver using chemical engines in an orbit around Earth towards the Sun is executed, lowering the spacecraft's orbit around the Sun (\hyperref[scenarios]{Table~\ref*{scenarios}}). Once it reaches a predefined solar distance of 0.6\,AU (black circle in \hyperref[sun-diver2d]{Fig.~\ref*{sun-diver2d}}), we assume that the sail is deployed, which initiates the second flight phase. From that point on, the spacecraft accelerates radially outward and therefore is pushed away from the Sun. We tested this method both for a transfer to Mars and to interstellar trajectories.

In addition to an increase of the terminal velocity this Sun diver approach offers another advantage. By adjusting the breaking velocity from the parking orbit around Earth the rapidity of the fall towards the Sun can be controlled. In \hyperref[sun-diver2d]{Fig.~\ref*{sun-diver2d}} we present the trajectory scatter for various values of ${\Delta v}_{\rm breaking}$. The $\Delta v$ was calculated considering the spacecrafts standard configuration with respect to the initial velocity of Earth's orbit ($v_{\rm Earth}=29.78\,{\rm km\,s}^{-1}$). Steeper dives require stronger breaking. By applying the correct amount of ${\Delta v}_{\rm breaking}$ or deploying the sail at the second intersection between the initial trajectory and the minimum distance every direction is possible inside Earth's orbital plane.

To implement the Sun diver maneuver some assumptions had to be made. To minimize the required $\Delta v$ for the breaking maneuver at Earth, the minimum distance around the Sun (the deployment altitude) must not be too low. Yet, in order to reduce the transfer time and to exploit the advantages of the first-in-then-out approach the minimum distance needs to be sufficiently low. Furthermore, for a single impulse maneuver in Earth's SOI, Mars needs to be located in the leading node of the two orbital planes upon arrival of the spacecraft. Multi-impulse maneuvers are also possible but have not been tested for this publication. With a multi-impulse maneuver the inclination change can be compensated and furthermore the range of possible trajectories around the Sun be increased.

\subsubsection{Inward Mars Transfer} \label{InwardMarsTransfer}

\begin{figure}[t]
	\centering
		\input{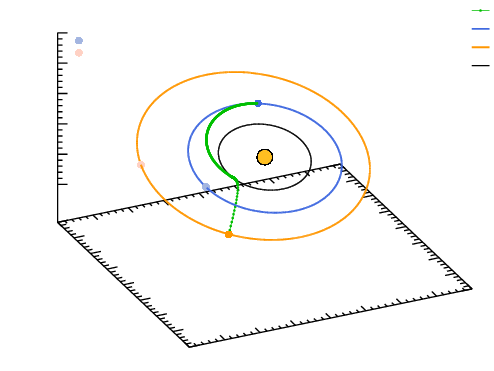}
	  \caption{Solar system showing Earth's and Mars' orbit around the Sun. The calculated trajectory for the inward Mars transfer reaches Mars in the leading node of both orbital planes.}
        \label{inwardMars2dSunSOI}
\end{figure}

\noindent
For the Mars transfer a random constellation between Earth and Mars was generated. As a boundary condition, however, we searched for trajectories from Earth that lead to encounters with Mars when Mars is in the ascending node of the two orbital planes. In these cases the spacecraft does not need to leave the ecliptic and we can consider the problem in a plane. To inject the spacecraft towards the Sun, a negative impulse of $\Delta v\approx4.5\,{\rm km\,s}^{-1}$ needs to be applied in the parking orbit around Earth. The minimum solar distance for sail deployment was again set to 0.6\,AU. Our numerical script contains two loops referring to the solar SOI as well as the martian SOI. In \hyperref[inwardMars2dSunSOI]{Fig.~\ref*{inwardMars2dSunSOI}} the solar system is shown with the additional orbit of Mars (orange) next to the orbit of Earth (blue). Both the position of Earth on arrival and the position of Mars at launch are shown faded. The required trajectory to reach Mars is presented in green.

\begin{figure}[t]
	\centering
		\input{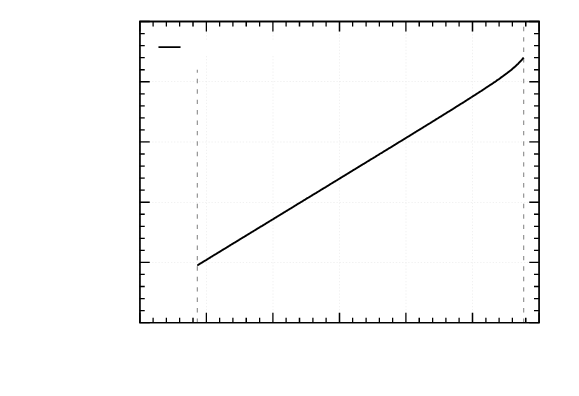}
	  \caption{Diagram of the spacecraft's velocity $v$ as a function of time $t$ inside the martian SOI for an inward transfer.}
        \label{vt3DiagramInMarsSOISun}
\end{figure}

Using the standard spacecraft configuration we calculate a total transfer time of 125.6\,d or 4.2 months (\hyperref[vt3DiagramInMarsSOISun]{Fig.~\ref*{vt3DiagramInMarsSOISun}}). When reaching the martian orbit the spacecraft has a velocity of $118.02\,{\rm km\,s}^{-1}$. Considering the entry velocity of $\approx117.85\,{\rm km\,s}^{-1}$, we find that it gained $\Delta v=0.17\,{\rm km\,s}^{-1}$ inside the martian SOI. Due to the high velocity the time spent in the martian SOI is just $\Delta t_{\mathrm{Mars}}\approx2.5\,{\rm hr}$. 

In summary, the sail was deployed after it reached the minimum solar distance of 0.6\,AU about 103\,d after launch. The maximum acceleration was determined as roughly $0.11\,{\rm m\,s^{-2}}$.

\subsubsection{Inward Interstellar Transfer}

\begin{figure}[t]
	\centering
		\input{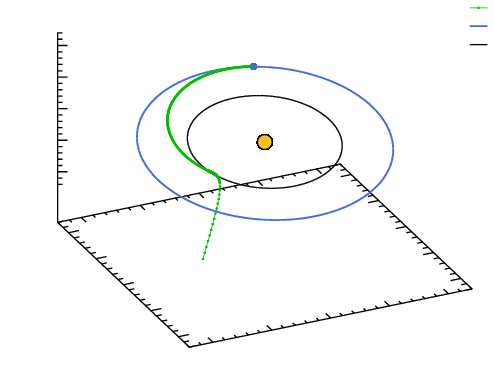}
	  \caption{Solar system showing Earth's orbit around the Sun. The calculated trajectory for the inward interstellar transfer escapes the solar system in a straight line.}
        \label{shotInInterstellar}
\end{figure}

\noindent
For the inward interstellar transfer the same approach was used as in subsection \hyperref[InwardMarsTransfer]{\ref*{InwardMarsTransfer}} for the Sun diver maneuver to Mars (\hyperref[scenarios]{Table~\ref*{scenarios}}). Starting in a solar orbit equally to the Earth's orbit the orbit of the spacecraft is lowered by initiating a negative impulse maneuver. By reaching the \texttt{minimum\_distance} the sail is deployed and the solar radiation pressure pushes the spacecraft away from the Sun. Yet, instead of stopping the calculating function when reaching the Mars SOI the trajectories are continued. As in subsection \hyperref[OutwardInterstellarTransfer]{\ref*{OutwardInterstellarTransfer}} the function then is stopped when the spacecraft reaches a distance of\\ $r=120\,{\rm AU}$. Again the trajectories showed a linear progression after performing the Sun diver maneuver in the close vicinity of the Sun (\hyperref[shotInInterstellar]{Fig.~\ref*{shotInInterstellar}}). 

As shown in \hyperref[atDiagramOutInterstellar]{Fig.~\ref*{atDiagramOutInterstellar}} a spacecraft with standard configuration on an inward interstellar transfer experiences a maximum acceleration of $0.11\,{\rm m\,s}^{-2}$ occurring after $103\,{\rm d}$. After the sail deployment the velocity increases again to its maximum of $148\,{\rm km\,s}^{-1}$ (\hyperref[vt2DiagramOutInterstellar]{Fig.~\ref*{vt2DiagramOutInterstellar}}). This maximum velocity is reached after about $300\,{\rm d}$. Flying with its maximum velocity the spacecraft reaches the boundary of the solar system after roughly $1520\,{\rm d}$ or $t_{\mathrm{end}}=4.16\,{\rm yr}$. 

Even closer solar approaches than those at 0.6\,AU that we tested are certainly possible. For context, the still active Parker Solar Probe has previously survived several hours interior to 0.1\,AU and as close as 0.086\,AU from the Sun \citep{2021PhRvL.127y5101K}.

\section{Conclusions}\label{Conclusions}

\noindent
We find that a 1\,kg solar sail with a cross-sectional area of $S=(100\,{\rm m})^2 = 10^4\,{\rm m}^2$ (radius of 56\,m) can be accelerated to a velocity of $v_{\mathrm{entry}}=65\,{\rm km\,s}^{-1}$ relative to Mars in a direct outward transfer from Earth (when Mars is near opposition). In this scenario, Mars is reached $26\,{\rm d}$ after launch. An initial Sun diver maneuver down to 0.6\,AU from the Sun increases the travel time and flyby velocities to Mars but allows launches from various Earth-Mars orbital constellations.

Our simulations of direct launches to an interstellar trajectory reveal terminal escape velocities of $v_{\mathrm{end}}=109\,{\rm km\,s}^{-1}$ and passage through the heliopause after 5.3\,yr. If instead the sail is first directed to the Sun, then a final velocity of $v_{\mathrm{end}}=148\,{\rm km\,s}^{-1}$ is obtained with a transfer time of $t_{\mathrm{end}}=4.2\,{\rm yr}$ to the heliopause. Details depend on how fast the sail is directed to the Sun (e.g. using conventional chemical vehicles) and how close its perihelion distance.

The trajectories and solar sail properties presented in this paper can be considered as a next step after the successful IKAROS mission to Venus, which had a comparable mass (0.5\,kg) but smaller sail area $196\,{\rm m}^2$. The key difference is in our assumption of aerographite as a sail material compared to the polyimide sheet of IKAROS, which had a mass of about $0.01\,{\rm kg\,m}^{-2}$ compared to $7.6 \times 10^{-7}\,{\rm kg\,m}^{-2}$ for aerographite.

An as of yet unresolved question about our concept is the deceleration of the payload if this is supposed to be a delivery and not a flyby mission. One possible scenario that we suggest for future studies is the deceleration using friction with the Martian atmosphere. Once dropped on the martian surface, radioactive contamination of the payload can help to recover even gram-sized payload on the Martian surface. The key challenges would, of course, be the structural integrity of the payload (not necessarily of the sail) during descent given the expected high acceleration and temperature. But the low gravity and the extended atmosphere of Mars actually yield a relatively wide deceleration corridor and moderate heating rates compared to Earth and Venus, making Mars a plausible target for aerocapture \citep{2023arXiv230810384P}.\\

\noindent
{\bf Acknowledgements}: R.~H. acknowledges support from the German Aerospace Agency (Deutsches Zentrum f\"ur Luft- und Raumfahrt) under PLATO Data Center grant 50OO1501.

\bibliographystyle{elsarticle-num}
\bibliography{main}

\end{document}